\documentclass[a4paper,11pt]{article}
\usepackage{styleBuding}

\usepackage[T1]{fontenc} 

\usepackage{slashed}
\usepackage[colorlinks=true,linktocpage=true]{hyperref}
\usepackage{cleveref}
\usepackage[numbers,sort&compress]{natbib}
\usepackage{mathtools}
\usepackage{xspace}

\usepackage{natbib}  

\usepackage{tikz}
\usepackage{subfigure}
\usepackage[section]{placeins}
\usepackage{fancyhdr}
\usepackage{array}
\usepackage{mathtools}
\usepackage{amssymb}
\usepackage{graphicx}
\usepackage{epsfig}
\usepackage{wrapfig}
\usepackage{graphics}
\usepackage{slashed}	
\usepackage{enumitem}
\usepackage{color}

\definecolor{SeaBlue}{rgb}{0.1,0.4,0.85}
\definecolor{DarkBlue}{rgb}{0.0,0.0,0.7}
\definecolor{NavyBlue}{rgb}{0.0,0.0,0.4}
\definecolor{Maroon}{rgb}{0.6,0.2,0.2}
\definecolor{SeaGreen}{rgb}{0.2,0.4,0.2}
\definecolor{Purple}{rgb}{0.7,0.1,0.6}
\definecolor{Red}{rgb}{0.8,0.2,0.2}
\definecolor{Black}{rgb}{0.0,0.0,0.0}
\definecolor{ZQS}{rgb}{0.01,0.227,0.451}
\definecolor{TS}{rgb}{0.588,0.31,0.05}

\usepackage[T1]{fontenc}
\usepackage{titlesec}
\titleformat{\section}
  { \color{NavyBlue} \normalfont \scshape}
  {\thesection}{1em}{\bf \MakeUppercase}
\titleformat{\subsection} { \color{NavyBlue}  \normalfont\scshape}{\thesubsection}{1em}{\bf }{}
\titleformat{\subsubsection}
  { \color{NavyBlue}  \normalfont\itshape}{\thesubsubsection }{1em}{}{}

\usetikzlibrary{decorations.pathmorphing}	
\usetikzlibrary{decorations.markings}
\tikzset{
    vector/.style={decorate, decoration={snake}, draw},
	provector/.style={decorate, decoration={snake,amplitude=2.5pt}, draw},
	antivector/.style={decorate, decoration={snake,amplitude=-2.5pt}, draw},
    fermion/.style={draw=black, postaction={decorate},
        decoration={markings,mark=at position .55 with {\arrow[draw=black]{>}}}},
    fermionbar/.style={draw=black, postaction={decorate},
        decoration={markings,mark=at position .55 with {\arrow[draw=black]{<}}}},
    fermionnoarrow/.style={draw=black},
    gluon/.style={decorate, draw=black,
        decoration={coil,amplitude=4pt, segment length=5pt}},
    scalar/.style={dashed,draw=black, postaction={decorate},
        decoration={markings,mark=at position .55 with {\arrow[draw=black]{>}}}},
    scalarbar/.style={dashed,draw=black, postaction={decorate},
        decoration={markings,mark=at position .55 with {\arrow[draw=black]{<}}}},
    scalarnoarrow/.style={dashed,draw=black},
    electron/.style={draw=black, postaction={decorate},
        decoration={markings,mark=at position .55 with {\arrow[draw=black]{>}}}},
	bigvector/.style={decorate, decoration={snake,amplitude=4pt}, draw},
}

\tikzstyle{block} = [draw, rectangle, 
    minimum height=3em, minimum width=6em]

\definecolor{cerulean}{rgb}{0., 0.52,0.65}

\newcolumntype{P}[1]{>{\raggedright\arraybackslash}p{#1}}

\usepackage{aas_macros}

\makeatletter
\def\l@subsubsection#1#2{}
\makeatother

\subheader{
\textrm{\begin{flushright}
APCTP-Pre2023-008 \\
\end{flushright}}}

\title{\boldmath Freeze-in bino dark matter in high scale supersymmetry}

\author[a, b]{Chengcheng Han,}
\author[c]{Peiwen Wu,}
\author[d,e]{Jin Min Yang,}
\author[f]{Mengchao Zhang}

\affiliation[a]{School of Physics, Sun Yat-Sen University, Guangzhou 510275, P. R. China}
\affiliation[b] {Asia Pacific Center for Theoretical Physics, Pohang 37673, Korea}
\affiliation[c]{School of Physics, Southeast University, Nanjing 211189,  P. R. China}
\affiliation[d]{CAS Key Laboratory of Theoretical Physics, Institute of Theoretical Physics, Chinese Academy of Sciences, Beijing 100190, P. R. China}
\affiliation[e]{School of Physics, Henan Normal University, Xinxiang 453007,  P. R. China}
\affiliation[f]{Department of Physics and Siyuan Laboratory, Jinan University, Guangzhou 510632, P. R. China }

\emailAdd{hanchch@mail.sysu.edu.cn}
\emailAdd{pwwu@seu.edu.cn}
\emailAdd{jmyang@itp.ac.cn}
\emailAdd{mczhang@jnu.edu.cn} 

\abstract{
We explore a scenario of high scale supersymmetry  where all  supersymmetric particles except gauginos stay at a high energy scale $M_{\rm SUSY}$ which is much larger than the reheating temperature $T_\text{RH}$. The dark matter is dominated by bino component with mass around the electroweak scale and the observed relic abundance is mainly generated by the freeze-in process during the early universe. Considering the various constraints, we identify two available scenarios in which the supersymmetric sector at an energy scale below $T_\text{RH}$ consists of: a) bino; b) bino and wino. 
Typically, for a bino mass around 0.1-1 TeV and a wino mass around 2 TeV, we find that $M_{\rm SUSY}$ should be around $10^{13-14}$ GeV with $T_\text{RH}$ around $10^{4-6}$  GeV.}

\begin{document}
\maketitle
\flushbottom


\section{Introduction  }
\label{sec:introduction}
Supersymmetry (SUSY) \cite{Golfand:1971iwegdsf,Volkov1973ykgwj,Wess1974sbnny,Salam1974jyfg,Wess1974adtukf,Ferrara1974wkxfdk} is a significant theoretical framework aiming at extending the Standard Model (SM), drawing inspiration from the pursuit of a quantum gravity theory, particularly within the context of superstring theory. In the field of phenomenology, SUSY not only provides a viable candidate for dark matter (DM) which plays a crucial role in the formation of large-scale structures in the universe, but also contributes to the renormalization group running of gauge couplings through the inclusion of additional particles near the electroweak scale. This property of SUSY facilitates the potential unification of the three fundamental forces at high energy scales. It has long been postulated that SUSY DM takes the form of Weakly Interacting Massive Particles (WIMPs) that can be probed through diverse experiments \cite{Steigman1985werfd, Jungman1996hrte, MARTIN1998wghfd, Feng2010fjshj, Cao:2012rz, Cao:2013gba, Han:2013usa, Han:2013kga, Han:2015lma, Han:2016xet, Han:2016gvr}. 
However, the absence of confirmed DM signals poses significant challenges to the standing of SUSY DM. The current LHC search results indicate that SUSY particles seem to be heavier than the electroweak (EW) scale \cite{ATL-PHYS-PUB-2023-005,CMSPublic-PhysicsResultsSUS}, thus challenging the WIMP paradigm of SUSY (for recent reviews on SUSY in light of current experiments, see, e.g., \cite{Baer:2020kwz,Wang:2022rfd,Yang:2022qyz}).

Given the current situation, in this study we consider an alternative scenario of SUSY DM in which gauginos are located at a low energy scale while all other SUSY partners exist at a significantly higher scale $M_{\rm SUSY}$. This scenario is a special case of the Split SUSY \cite{Giudice2005fuyk,Arkani-Hamed2005htrewh,Arkani-Hamed2005ajdkud,Wells2005wmghd} where higgsinos are also taken to be a similar scale as sfermions. One should note that the Higgs sector in this scenario is fine-tuned~\cite{Hall:2009nd, Giudice:2011cg, Ibe:2012hu, Ibanez:2013gf, Hall:2014vga,  Ellis:2017erg, Liu:2005qic, Liu:2005rs} and it might be a consequence of the anthropic principle. However, in this work we will assume that SUSY still provides a candidate of DM and we will specifically consider the Minimal Supersymmetric Standard Model (MSSM). Since the measurement of gamma-ray from the MAGIC \cite{MAGIC:2022acl}  has strongly constrained the possibility of wino DM\footnote{There is still viable parameter space for wino dark matter assuming core profile of the DM.},  the only viable DM candidate in the MSSM is bino. However, it is widely known that pure bino DM is typically overabundant from the freeze-out mechanism \cite{Kolb:1990vq} due to its weak coupling with the visible sector \cite{Olive:1989jg,Griest:1989zh}. Alternatively, a bino particle with a rather weak coupling may serve as a suitable candidate for Feebly Interacting Massive Particle (FIMP) DM with a correct relic abundance generated via the freeze-in mechanism \cite{Hall:2009bx}, with assumptions that the reheating process solely occurs in the Standard Model (SM) sector and the reheating temperature $T_\text{RH}$ is lower than the SUSY scale $M_\text{SUSY}$ .

In this work we study the possibility that the bino DM in MSSM is generated via the freeze-in process during the early universe. We assume that all MSSM particles except gauginos share similar mass $M_{\rm SUSY}$ which is much higher than the reheating temperature $T_\text{RH}$ of the universe. To generate enough relic abundance of bino dark matter, we always require the bino mass lower than the reheating temperature. While for the mass of wino or gluino, they could be either higher or lower than the reheating temperate $T_\text{RH}$ depending on the different scenarios we consider. 

The paper is organized as follows. In Section \ref{sec-scenarios} we present the model set up. In Section \ref{sec-overview} we first overview the physics related to dark matter and then study the dominate channels for bino freeze-in production. In Section \ref{sec-numerical-results} we give the numerical results and discuss the experimental limits on the model parameter space relevant for our scenarios. We draw the conclusions in Section \ref{sec-conclusion} and leave  the calculation details in Appendices.

\section{Model of heavy supersymmetry  }
\label{sec-scenarios}
Since we are considering a scenario of high scale supersymmetry in which only gauginos are at low energy scale, the relevant Lagrangian terms are
\small 
\begin{eqnarray} 
\mathcal{L} &\supset& - \sum_{f=q,l} M^2_{\tilde{f}} \tilde{f}^\ast \tilde{f} + \bigg[   \bigg( \sum_{A=1,2,3} - \frac{1}{2}  M_A \tilde{V}^{A,a} \tilde{V}^{A,a} \bigg)  -  \mu  \tilde{H}_u \cdot \tilde{H}_d  + b \mu H_u \cdot H_d + h.c. \bigg]   \nonumber\\
& & - \sum_{A=1,2} \sqrt{2} g_A  \bigg[ {H}^{\ast}_{u} \bigg(T^{A,a} \tilde{V}^{A,a} \bigg)\tilde{H}_{u} + {H}^{\ast}_{d} \bigg(T^{A,a} \tilde{V}^{A,a}\bigg) \tilde{H}_{d} + h.c. \bigg] \nonumber\\
& & - \sum_{A=1,2,3} \sqrt{2} g_A  \bigg[ \sum_{f=q,l}  \tilde{f}^{\ast}  \bigg(T^{A,a} \tilde{V}^{A,a}\bigg) f + h.c. \bigg] \nonumber\\
& &- (M^2_{H_u}+|\mu|^2)H^\ast_u H_u - (M^2_{H_d}+|\mu|^2) H^\ast_d H_d ~,
\label{eq-L-master}
\end{eqnarray}
\normalsize 
where $A=1,2,3$ correspond to the SM gauge group $\rm  U(1)_Y, SU(2)_L, SU(3)_C$, respectively, and $a$ denotes the corresponding indices in adjoint representation of group $A$. Fields $\tilde{V}^{A,a}, \tilde{H}_{u}, \tilde{H}_{d}, \tilde{f}$ are the superpartners of the SM vector gauge bosons $V^{A,a}=B, W^{1\sim 3}, G^{1\sim 8}$, scalar doublets $H_u, H_d$ and fermions $f$. The fields $H_u$, $H_d$, $\tilde H_u$, $\tilde H_d$ are defined as
\begin{eqnarray} 
H_u &=& \left(\begin{array}{c} H_u^+ \\ H_u^0  \end{array}\right) ,\quad \tilde{H}_u = \left(\begin{array}{c} \tilde{H}_u^+ \\ \tilde{H}_u^0  \end{array}\right), \quad 
H_d = \left(\begin{array}{c} H_d^0 \\ H_d^-  \end{array}\right) , \quad \tilde{H}_d = \left(\begin{array}{c} \tilde{H}_d^0 \\ \tilde{H}_d^-  \end{array}\right).
 \end{eqnarray}
For the Higgs sector, we need a SM-like Higgs boson $H_\text{SM}$ near the electroweak scale \cite{ATLAS:2012yve,CMS:2012qbp}. This is obtained from the mixing between the two Higgs doublets $H_u$ and $H_d$ in the MSSM: 
\small 
 \begin{eqnarray} 
H_u = \left(\begin{array}{c} H_u^+ \\ H_u^0  \end{array}\right) 
&=& \sin\beta \, H_{\rm SM} + \cos\beta \, H_{\rm NP} =
 \sin\beta \left(\begin{array}{c} G_{\rm SM}^+ \\ H_{\rm SM}^0  \end{array}\right)
+\cos\beta \left(\begin{array}{c} H_{\rm NP}^+ \\ H_{\rm NP}^0  \end{array}\right)~,   \\
(-i \sigma^2)H^\ast_d 
= \left(\begin{array}{c} -(H_d^-)^* \\ (H_d^0)^*  \end{array}\right)
&=& \cos\beta \, H_{\rm SM}- \sin\beta \,  H_{\rm NP} =
 \cos\beta \left(\begin{array}{c} G_{\rm SM}^+ \\ H_{\rm SM}^0  \end{array}\right)
-\sin\beta \left(\begin{array}{c} H_{\rm NP}^+ \\ H_{\rm NP}^0  \end{array}\right)~,
 \end{eqnarray}
 \normalsize
where $\sigma^2$ is the second Pauli matrix, and $\tan\beta = \langle H^0_u \rangle / \langle H^0_d \rangle$ with $ \langle H^0_u \rangle$ and $\langle H^0_d \rangle$ being the vacuum expectation values (VEVs). Such mixings can be realized by properly choosing Higgs mass parameters $\mu, M_{H_u}, M_{H_d}$, and $b$.
The subscription "NP" in $H_\text{NP}$ denotes the new physics (NP) Higgs doublet in the MSSM accompanying the SM one\footnote{Note that in order not to increase the complexity of notation, we don't further perform the expansion of the complex but electrically neutral scalars $H_{\rm SM}^0, H_{\rm NP}^0$ into real and imaginary parts. However, one needs to beware that $G_{\rm SM}^\pm, H_{\rm SM}^0$ contain the Goldstone boson modes to be absorbed into vector gauge bosons $W^\pm, Z^0$ after the electroweak symmetry breaking (EWSB).}.  Since the mass parameters $M_{H_u}$,  $M_{H_d}$, $b$, $\mu$ are all much larger than the electroweak scale, a tuning of these parameters are needed to get a light Higgs at electroweak scale~\cite{Hall:2009nd, Giudice:2011cg, Ibanez:2013gf, Hall:2014vga, Ellis:2017erg}.  We need also match the Higgs self-coupling to be the SUSY value at the scale of $M_{\rm SUSY}$,
\small 
\begin{eqnarray} 
\lambda(M_{\rm SUSY}) =\frac{{g^\prime_1}^2+g_2^2}{4} \cos^2 2\beta~.
\end{eqnarray}
\normalsize 
Note that the Higgs self-coupling $\lambda$ becomes very small at high energy scale due to the RGE running, and thus the $\beta$ value should get close to $\pi/4$ and $\tan\beta \approx 1$. We will fix $\tan\beta = 1$ as the benchmark parameter throughout this work for simplicity.


Generally, when considering physical processes at temperature $T \ll M_\text{SUSY}$,  we can integrate out the heavy mediators with mass $\mu,M_{\tilde{f}} \sim M_\text{SUSY} \gg T_\text{RH}$ and get the following effective operators at the level of dimension 5 and 6, respectively,
\begin{eqnarray} 
&&{\text{dimension-5:}} \quad \propto \quad \frac{1}{\mu} \, | H_{\rm SM} |^2   ( \tilde{B} \tilde{B}, \tilde{B} \tilde{W} )~,  \\
&&{\text{dimension-6:}} \quad \propto \quad \frac{1}{M^2_{\tilde{f}}} ( f^{\dagger} \tilde{B}^\dagger ) (f \tilde{B}, f \tilde{W}, f \tilde{G})~.
\label{Leff_overview}
\end{eqnarray} 
Since we assume the mass parameters of higgsinos $\mu$ and sfermions $M_{\tilde{f}}$ around $M_\text{SUSY}$, the dominant process would be from the dimension-5 (dim-5) operators. Nevertheless we also present the processes related to dim-6 operators for completeness.

We acknowledge that a majority of the significant processes are evaluated at energy scales considerably beneath $M_\text{SUSY}$. The recommended approach entails initiating the integration procedure for the massive particle to derive the effective operators of dimension 5 and 6, along with their corresponding Wilson coefficients, within the realm of $M_\text{SUSY}$. Subsequently, the computation of these Wilson coefficients at the pertinent scale is achieved by employing the Renormalization Group Equations to track the evolution of the operators. Notably, there exists a potential correction to the primary outcome, potentially on the order of $O(1)$, yet the fundamental framework remains robust. We leave the investigation of this effect for future study.

\section{Freeze-in bino dark matter in MSSM}
\label{sec-overview}

\subsection{Particle spectrum}
Despite the existence of new Higgs bosons and many supersymmetric partners of the SM particles, the MSSM particle spectrum we consider in this work consist of two sectors distinguished by their characteristic mass scales. Although not making significant difference for the mass spectrum structure before and after EWSB, we take the pre-EWSB case as an illustration.
\begin{itemize}
\item {\bf{Heavy}} sector, {\bf{inactive}} after cosmological reheating

Mass: $M\sim M_\text{SUSY} \gg T_{RH}$

\begin{itemize}
\item  Higgs bosons not in SM: $H^0_\text{NP}$, $A$, $H^\pm_\text{NP}$
\item Sfermions $\tilde{f}$
\item Higgsinos $\tilde{H}_u,\tilde{H}_d$
\end{itemize}

\item {\bf{Light}} sector, {\bf{active}} after cosmological reheating

Mass: $M \sim \mathcal{O}(1) \, \text{TeV} \ll M_\text{SUSY} $
\begin{itemize}
\item SM particles
\item Bino  $\tilde{B}$, consisting cosmological DM with mass $M_1 < T_{RH}$
\item Winos $\tilde{W}$, with mass $M_2$
\item Gluinos $\tilde{G}$, with mass $M_3$
\end{itemize}
\end{itemize}
In the above we utilized gauge eigenstates for description, since $\tilde{B},\tilde{W}$ do not mix with higgsinos $\tilde{H}_u, \tilde{H}_d$ before EWSB when the SM Higgs $H_\text{SM}$ has not acquired the VEV.


\subsection{Bino production from freeze-in mechanism  }
\label{bino-freezein}
\begin{figure}[h]
  \centering
  \includegraphics[width=6.5cm]{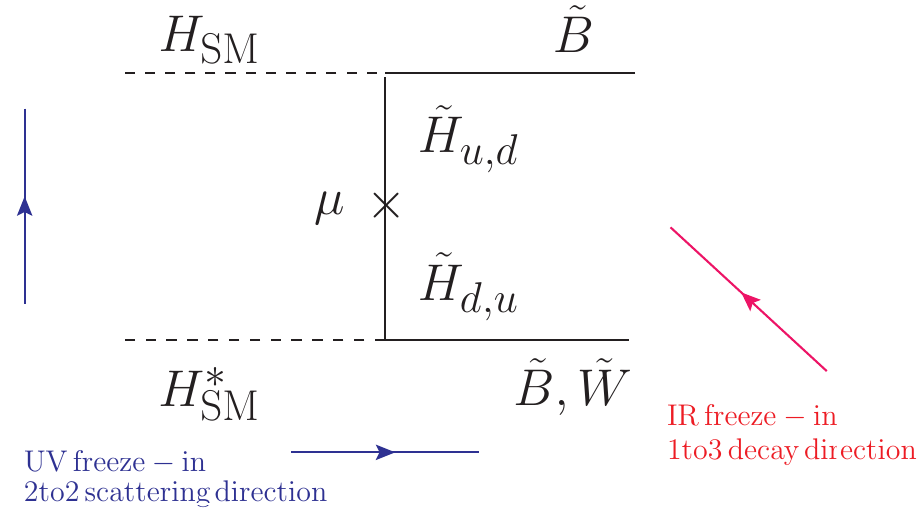} \hspace{1 cm}
   \includegraphics[width=6.5cm]{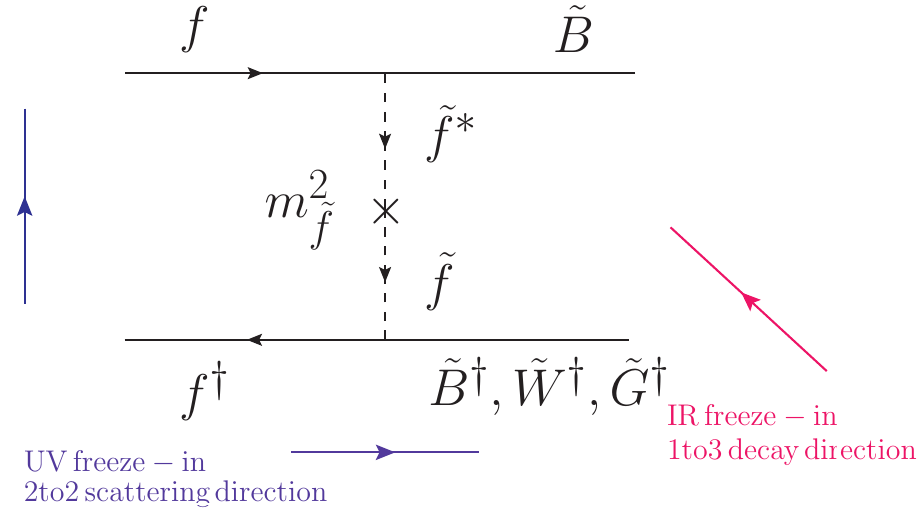}
  \caption{Schematic plots for interactions of DM composed of pure $\tilde{B}$ with SM after cosmological reheating considered in this work, which would induce dimension-5 (left) and dimension-6 (right) effective operators. The SM Higgs $H_\text{SM}$ originates from the mixing between MSSM Higgs doublets $H_u, H_d$.
  Colored lines indicate the direction of freeze-in production when applicable. Additional Hermitian conjugated processes also exist when the amplitudes are complex. See more discussions in the main texts.}
  \label{fig-schematic-process}
\end{figure}

In the early stage of universe before EWSB when the gaugino states $\tilde{B},\tilde{W}$ do not mix with higgsinos $\tilde{H}_u, \tilde{H}_d$, pure $\tilde{B}$ acting as DM can only interact with SM via mediators with heavy mass near the scale $M_\text{SUSY}$,  as shown in Fig.~\ref{fig-schematic-process}. Due to the suppressed interacting strength, the cosmological production  of bino DM in our scenario proceed via the freeze-in mechanism.
In the follows we consider the contributions to bino DM production from several typical processes\footnote{After electroweak phase transition occurs and $H_\text{SM}$ acquires VEV, the top and bottom vertex in the left panel of Fig.\ref{fig-schematic-process} imply the mixing between $\tilde{B}, \tilde{W}$ and $\tilde{H}_u,\tilde{H}_d$, resulting in the mass eigenstates of electrically neutral neutralinos $\tilde{\chi}^0_{1,2,3,4}$ and charged $\tilde{\chi}^\pm_{1,2}$ (see discussions in Section \ref{sec:BBN}).}.

\subsection{Case I: bino freeze-in from $HH^\ast \to \tilde{B}\tilde{B}$} 
\label{Case-I}

This case corresponds to the left panel of Fig.~\ref{fig-schematic-process} but without winos $\tilde{W}$. After integrating out the heavy higgsinos, the relevant dim-5 effective interaction is given by (the details are given in Appendix \ref{Appendix-A})
\begin{eqnarray} 
\mathcal{L}^{\text{eff}}_{HH^\ast \to \tilde{B}\tilde{B}} 
= \frac{2 g_1^2 \, Y_H^2}{\mu} \sin\beta  \cos\beta  ( | H_{\rm SM} |^2 )  ( \tilde{B} \tilde{B} + \tilde{B}^{\dagger} \tilde{B}^{\dagger} )~,
\label{Leff_HHBB}
\end{eqnarray} 
where $| H_{\rm SM} |^2 = G_{\rm SM}^+ (G_{\rm SM}^+)^\ast + (H_{\rm SM}^0)(H_{\rm SM}^0)^*$. In the subscription $HH^\ast \to \tilde{B}\tilde{B}$ on the left side (and hereafter when not causing any confusion), we denote $H_\text{SM}$ as $H$ to simplify the notation, and all fields in the initial and final states of the process should be understood in the sense of physical particles\footnote{Discussion on the naming convention of particles, states and filed can be found in, e.g. \cite{Dreiner:2008tw}.}.
With more details given in Appendix \ref{Appendix-B},  Eq.(\ref{Leff_HHBB}) would induce the Boltzmann equation of the bino number density: 
\begin{eqnarray}
\frac{d}{d t} n_{\tilde{B}} + 3 \mathcal{H} n_{\tilde{B}} = \textbf{C}_{HH^\ast \to \tilde{B}\tilde{B}} \approx \frac{ g_1^4  }{ 4 }  \frac{1}{ \pi^5} \frac{ \, \sin^2\beta  \cos^2\beta }{\mu^2 }  T^6~.
\end{eqnarray}
The above equation can be modified to a differential equation about bino yield $Y_{\tilde{B}} = n_{\tilde{B}}/S  $ ($S$ is the entropy density) and temperature $T$:
\begin{eqnarray}
&& \frac{d Y_{HH^\ast \to \tilde{B}\tilde{B}}(T)}{d T} = - \frac{\textbf{C}_{HH^\ast \to \tilde{B}\tilde{B}}}{ST\mathcal{H}} \nonumber \\
&&\approx -(1.25 \times 10^{-3}) \times M_{\text{Pl}} \frac{\textbf{C}_{HH^\ast \to \tilde{B}\tilde{B}}}{ T^6 }   \nonumber 
\approx - (1\times 10^{-6}) \times  M_{\text{Pl}} \  g_1^4 \frac{ \, \sin^2\beta  \cos^2\beta }{\mu^2 }~,
\end{eqnarray}
where $ M_{\text{Pl}} \approx 1.22\times 10^{19}$ GeV is the Planck mass, 
$S = 2\pi^2 g_\ast T^3 / 45$, and Hubble expansion rate $\mathcal{H} \approx 1.66 \sqrt{g_\ast} T^2 / M_{\text{Pl}} $ with $g_\ast = 106.75$ before EWSB. 
Performing a simple integral from reheating temperature, it can be found that the final yield of $\tilde{B}$ depends on the reheating temperature $T_{\text{RH}}$ which corresponds to the Ultraviolet (UV) freeze-in scenario \cite{Hall:2009bx,Elahi:2014fsa}: 
\begin{eqnarray}
Y_{HH^\ast \to \tilde{B}\tilde{B}}(\infty)  \approx (1\times 10^{-6}) \times  M_{\text{Pl}} \ g_1^4  \frac{ \, \sin^2\beta  \cos^2\beta }{\mu^2 }  \ T_{\text{RH}}~,
\end{eqnarray}
and the corresponding current relic abundance is given by
\begin{eqnarray}
\left(\Omega_{\tilde{B}} h^2 \right)_{HH^\ast \to \tilde{B}\tilde{B}} = M_1 \ \frac{Y_{HH^\ast \to \tilde{B}\tilde{B}}(\infty) S_0 }{ \rho_{cr}  }  \approx Y_{HH^\ast \to \tilde{B}\tilde{B}}(\infty) \left( \frac{M_1}{\text{TeV}} \right) \times (2.72\times 10^{11}) ~. \nonumber \\
\end{eqnarray}

\subsection{Case II: fermion scattering process $f \bar{f} \to \tilde{B}\tilde{B}$}
 \label{fermion}
After integrating out sfermions with heavy mass $M_{\tilde{q},\tilde{l}}\sim M_\text{SUSY}$ in the right panel of Fig.\ref{fig-schematic-process}, the effective interactions between SM fermion pair and $\tilde{B}$ pair have the following form at dimention 6 (for more details, see Appendix \ref{Appendix-C}):
\begin{eqnarray} 
\mathcal{L}^{\text{eff}}_{f\bar{f} \to \tilde{B}\tilde{B}} =  \sum_{f=q,l} \frac{ (\sqrt{2} g_1 Y_f)(\sqrt{2} g_1 Y_f)}{M^2_{\tilde{f}}} ( f^{\dagger} \tilde{B}^\dagger ) (f \tilde{B}) ~,
\end{eqnarray} 
where for simplicity, we consider an universal mass for all the fermions, i.e. $M_{\tilde{f}} \equiv M_{\tilde{q}} = M_{\tilde{l}}$.

Thus the Boltzmann equation is 
\begin{eqnarray}
&&\frac{d Y_{f\bar{f} \to \tilde{B}\tilde{B}} (T)}{d T}  = - \frac{\textbf{C}_{f\bar{f} \to \tilde{B}\tilde{B}}}{ST\mathcal{H}} \nonumber \\ 
&& \approx -(1.25 \times 10^{-3}) \times M_{\text{Pl}} \frac{\textbf{C}_{f\bar{f} \to \tilde{B}\tilde{B}} }{ T^6 } 
\approx - (8.6\times 10^{-5}) \times  M_{\text{Pl}} \frac{g_1^4}{M^4_{\tilde{f}}} T^2~,
\end{eqnarray}
and correspondingly,
\begin{eqnarray}
Y_{f\bar{f} \to \tilde{B}\tilde{B}}(\infty)  &\approx& (4.7\times 10^{-7}) \times \frac{M_{\text{Pl}} }{M^4_{\tilde{f}}} \ T^3_{\text{RH}} ~, \\ 
\left( \Omega_{\tilde{B}} h^2 \right)_{f\bar{f} \to \tilde{B}\tilde{B}} &=& M_1 \ \frac{Y_{f\bar{f} \to \tilde{B}\tilde{B}}(\infty) S_0 }{ \rho_{cr}  }  \approx Y_{f\bar{f} \to \tilde{B}\tilde{B}}(\infty) \left( \frac{M_1}{\text{TeV}} \right) \times (2.72\times 10^{11})  ~.
\end{eqnarray}

\subsection{Case III: gluino/wino scattering or decay processes}
\label{Sec-case-III}
As indicated by blue colored arrows in Fig.~\ref{fig-schematic-process}, the $2\to 2$ scattering processes consist of two ways of generating bino DM when combining $\rm U(1)_Y$ with $\rm SU(2)_L$ or $\rm SU(3)_C$ interactions, related by the cross symmetry.
Moreover, we can also have the red colored arrow indicating $1\to3$ ($1\to2$) decay processes generating binos before (after) EWSB
when the cosmological temperature drops below the scale of $M_2$ or $M_3$ (equivalently, when the age of the universe reach the lifetime of $\tilde{W}$ and $\tilde{G}$).

Similar to the previous two cases, integrating out heavy higgsino and sfermions would generate the following dim-5 and dim-6 effective operators:
\begin{eqnarray} 
\mathcal{L}^{\text{eff}}_\text{case-III} = & & \bigg\{ - \sum_{b=1}^{3} \frac{ (\sqrt{2} g_1 Y_H) (\sqrt{2} g_2  ) }{\mu} \sin\beta \cos\beta (H^\ast  \frac{1}{2}\sigma^b  H) (\tilde{B}\tilde{W}^b)  \nonumber\\
& & + \sum_{f= u_L, d_L, e_L, \nu} \quad \sum_{b=1}^{3} \frac{(\sqrt{2} g_1 Y_f) (\sqrt{2} g_2) }{M^2_{\tilde{f}}} ( f^{\dagger} \tilde{B}^\dagger ) (\frac{1}{2}\sigma^b   f \tilde{W}^b) \nonumber \\
&  & +  \sum_{f=u_L,d_L,{u^\dagger_R},{d^\dagger_R}} \quad \sum_{a=1}^{8} \frac{ (\sqrt{2} g_1 Y_f) (\sqrt{2} g_3 ) }{M^2_{\tilde{f}}} ( f^{\dagger} \tilde{B}^\dagger ) (\frac{1}{2} \lambda^{a}  f \tilde{G}^a)  \bigg\} + h.c. ~.
\end{eqnarray} 
Note that the index $f$ in the second line includes only $\rm SU(2)_L$ doublets, while the index $f$ in the third line includes only quarks. To highlight the difference, we use index $a$ and $b$ to denote generators of $\rm SU(3)_C$ and $\rm SU(2)_L$ interactions, respectively. Correspondingly, $\lambda^{a}$ and $\sigma^b$ are Gell-Mann and Pauli matries, respectively.

In the following, we consider the contributions to the bino DM production from $2\to2$ scattering and $1\to3$ decay separately, while leaving the effects of $1\to2$ decay appearing after EWSB in Section \ref{sec:BBN}.

\subsubsection{Case III A: $2\to2$ scattering involving gluino/wino}
\label{Gaugino_Scat}

With more detailed given in Appendix \ref{Appendix-D}, the collision terms in the Boltzmann equation for dim-5 and dim-6 operators are approximated as (ignoring the masses of all external particles)
\small 
\begin{eqnarray}
\textbf{C}_{\rm dim-5}  & = &  \frac{T}{ 2048 \pi^6 } \int^{\infty}_{4M_1^2} ds \  ( s-4M_1^2 )^{1/2} K_1( {\sqrt{s}}/{T})  \sum_{\rm internal \, d.o.f}  \int\, d\Omega  \nonumber\\
& & \times  \bigg(  |\mathcal{M}|^2_{HH^\ast \to \tilde{B}\tilde{B}} +  |\mathcal{M}|^2_{HH^\ast \to \tilde{B}\tilde{W}} + N_\text{conj} |\mathcal{M}|^2_{\tilde{W} H \to \tilde{B} H}\bigg)  \nonumber\\
& & = \bigg( \frac{ 1 }{ 4 }g_1^4  + \frac{3 }{ 2 } g_1^2 g_2^2  \bigg)  \frac{1}{ \pi^5}  \frac{ \, \sin^2\beta  \cos^2\beta }{\mu^2 }   T^6~,\\
\textbf{C}_{\rm dim-6}  & = &  \frac{T}{ 2048 \pi^6 } \int^{\infty}_{4M_1^2} ds \  ( s-4M_1^2 )^{1/2} K_1( {\sqrt{s}}/{T})   \sum_{\rm internal \, d.o.f}  \int\, d\Omega   \nonumber\\
& &\times \bigg(|\mathcal{M}|^2_{f\bar{f} \to \tilde{B}\tilde{B}}+ |\mathcal{M}|^2_{f\bar{f} \to \tilde{B}\tilde{W}} + N_\text{conj} |\mathcal{M}|^2_{ \tilde{W} f \to \tilde{B} f } +  |\mathcal{M}|^2_{f\bar{f} \to \tilde{B}\tilde{G}} + N_\text{conj} |\mathcal{M}|^2_{ \tilde{G} f \to \tilde{B} f }  \bigg)  \nonumber\\
& =&  ( \frac{190}{9}g^4_1 + 30 g_1^2 g_2^2 + \frac{440}{3} g_1^2 g_3^2  )  \frac{1}{\pi^5 }  \frac{1}{M^4_{\tilde{f}}} T^8~,
\end{eqnarray}
\normalsize 
where $N_\text{conj}=2$ denotes the effects of conjugated process.

\subsubsection{Case III B: decay of gluino/wino}
\label{Gaugino_decay}

Following the method in ~\cite{Hall:2009bx} with $f_{\tilde{G}}$ and $f_{\tilde{W}}$ approximated by $e^{-E_{\tilde{G}}/T}$ and $e^{-E_{\tilde{W}}/T}$, the Boltzmann equation of freeze-in production for the $1\to3$ decay processes is 
\begin{eqnarray}
&&\frac{d}{d t} n_{\tilde{B}} + 3 \mathcal{H} n_{\tilde{B}} = \textbf{C} \nonumber\\
&&\approx  \frac{ g_{\tilde{G}} M_3^2  }{2\pi^2} T K_1(\frac{M_3}{T}) \Gamma_{\tilde{G} \to f\bar{f} \tilde{B} }
+ \frac{ g_{\tilde{W}} M_2^2  }{2\pi^2} T K_1(\frac{M_2}{T}) ( \Gamma_{\tilde{W} \to f\bar{f} \tilde{B} } + \Gamma_{\tilde{W} \to H H^\ast \tilde{B} }  )~,
\end{eqnarray}
where $g_{\tilde{G}} = 16$ and $g_{\tilde{W}} = 6$ are the internal d.o.f. of $\tilde{G}$ and $\tilde{W}$, respectively. The expressions of decay width involved in the above results are listed in Appendix \ref{Appendix-E}. Changing variables to yield $Y_{\tilde{B}}$ and temperature $T$, we then integrate over temperature evolution to obtain the final yield. If reheating temperature $T_\text{RH}$ is much larger than $M_2$ and $M_3$, then the final yield from $1\to3$ decay can be approximated by 
\small
\begin{eqnarray}
&& Y^{\rm 1\to3}_{\tilde{B}}(\infty) \approx \int^{T_{\text{RH}}}_{T_{min}} \frac{\textbf{C}}{ST\mathcal{H}} dT  \nonumber \\ 
&&\approx  (3\times10^{-4})\times M_{\text{Pl}} \left( \frac{ 1  }{M^2_3} g_{\tilde{G}} \Gamma_{\tilde{G} \to f\bar{f} \tilde{B} } +  \frac{ 1  }{M^2_2} g_{\tilde{W}} \Gamma_{\tilde{W} \to f\bar{f} \tilde{B} } + \frac{ 1  }{M^2_2} g_{\tilde{W}} \Gamma_{\tilde{W} \to H H^\ast \tilde{B} }    \right)~.
\end{eqnarray}
\normalsize 
It is worth pointing out that the above result is not sensitive to $T_{\text{RH}}$. Taking a low reheating temperature $T_{\text{RH}} = 1.1\,M_3$ as an example, increasing the value of $T_{\text{RH}}$ does no modify the result significantly.

In addition to the $1\to3$ decay, we should also note that wino $\tilde{W}$ with mass $M_2 < T_\text{RH}$ keeps staying in the thermal bath until reaching its freeze-out moment yielding a relic wino number density, which would later convert to the equal amount of bino number density $n_{\tilde{B}}$ via $1\to2$ decay $\tilde{W}\to\tilde{B}+h$ after EWSB occurs. Depending on the bino mass $M_1$, this freeze-out component would also contribute to the total bino DM abundance in today's epoch. We checked that with wino mass $M_2=2$ TeV, the $1\to2$ decay contribution of $Y^{1\to2}_{\tilde{B}}$ to final bino yield is around $25\%$ ($1\%$) on the percentage level for $M_1=$ 1(0.1) TeV \cite{Beneke:2020vff}, thus not affecting the freeze-in domination scenario of this work. We properly include the wino freeze-out contribution in our results. There is also contribution from gluino late time decay. However, to avoid the constraints from BBN, we have to set the gluino mass higher than the $T_\text{RH}$, thus we do not include its contribution here.



\section{Numerical results and discussions}\label{sec-numerical-results}



\begin{figure}[ht]
  \centering
  \includegraphics[width=8cm]{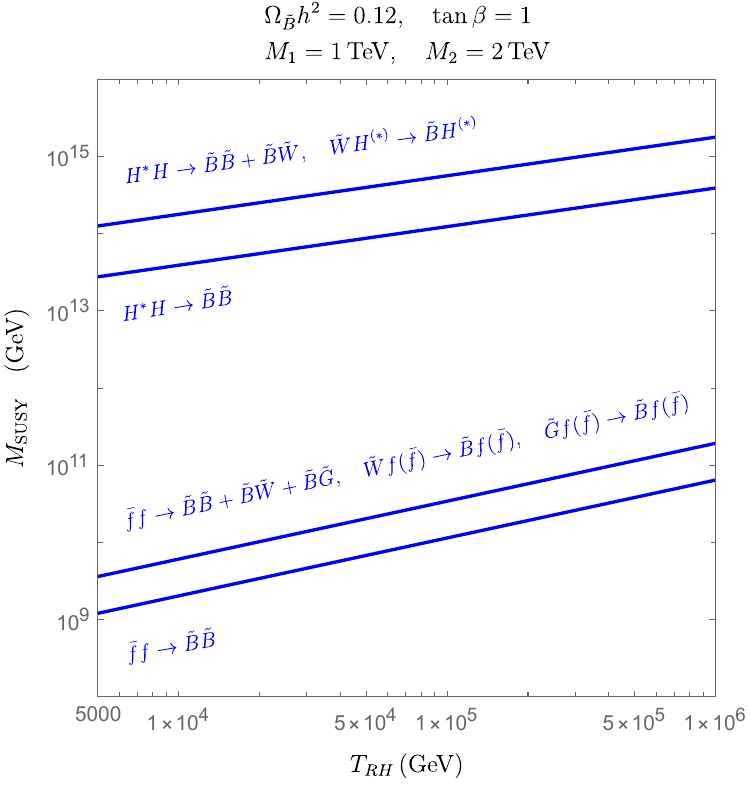} \hspace{1 cm}
  \caption{Values of $\mu$ and $M_{\tilde{f}}$ to produce the observed DM abundance via the UV freeze-in processes. See more discussions in the main texts.}
  \label{fig-02}
\end{figure}

In Fig.~\ref{fig-02} we show the required scales of $\mu$ ($M_{\tilde{f}}$) for dim-5(6) operators with various $T_\text{RH}$ to produce the observed bino DM relic abundance. The upper (lower) two lines correspond to dim-5 (6) operators. We can see that due to the more suppression of dim-6 operators, the needed  $M_{\tilde{f}}$ are generally $\mathcal{O}(10^{-4})$ smaller than $\mu$ in the dim-5 case. If we assume $\mathcal{O}(\mu)\approx\mathcal{O}(M_{\tilde{f}})$, in order not to overclose the Universe, the dim-6 contributions would be completely negligible. 

From Fig.~\ref{fig-02},  we can see that  for the case $M_{\tilde B} < T_{RH} \ll M_{\tilde W}$, the dominant production of bino dark matter is from the process $H H^* \rightarrow \tilde B \tilde B$ from the dim-5 operator. Generally, $M_{\rm SUSY}$ should be around $10^{13-14}$ GeV for $T_{RH} < 10^6$ GeV. Since the final relic abundance is proportional to  ${T_{RH}}/{\mu^2}$, the $M_{\rm SUSY}$ could continue increasing if the reheating temperature $T_\text{RH}$ becomes higher. Note that this is similar to the model of Higgs portal to fermion dark matter which are studied in \cite{Ikemoto:2022qxy}, with which we find our result are consistent. We emphasize that our model is motivated by a more complete framework and \cite{Ikemoto:2022qxy} falls into one of cases we consider. Moreover, 
For the case $M_{\tilde B},  M_{\tilde W}  < T_{RH} $, we find the wino-included process can largely enhance the annihilate rate and a higher scale is needed to satisfy the relic abundance. In this case, $M_{\rm SUSY}$ should be around $10^{14-15}$ GeV for $T_{RH} < 10^6$ GeV. 

Notice that if the gluino is in the thermal equilibrium with SM in the early universe and the sfermions mediating the gluino decay are heavier than $10^9$ GeV, the lifetime of the gluino could be longer than the age of the Universe when the big bang nucleosynthesis (BBN) happens, leading to energy injection into the cosmic plasma and altering the BBN profile. In all cases considered in this work we find $M_{\rm SUSY}$ is much larger than $10^9$ GeV, therefore we always need $M_{\tilde G} \gg T_{RH}$ to avoid the limit from BBN~\cite{Arvanitaki:2005fa}. More discussions on BBN limits are given in \ref{sec:BBN}.

\begin{figure}[ht]
  \centering
   \includegraphics[width=8cm]{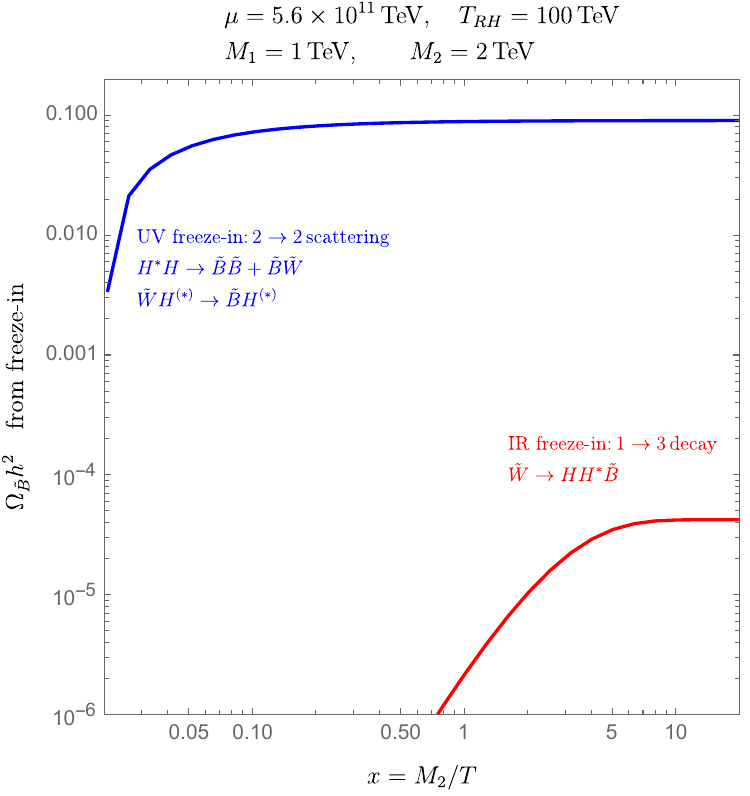}
  \caption{Comparison between UV freeze-in and IR freeze-in. Note the difference between temperatures indicated by $x=M_2/T$ producing the correct relic density of bino DM. }
  \label{fig-03}
\end{figure}

In Fig.~\ref{fig-03} we show the comparison of final contributions and intermediate profile of UV and IR freeze-in processes to the bino DM relic abundance. It can be clearly seen that the IR freeze-in final yields from  wino 3-body decays are negligible compared to that of UV freeze-in processes generated by $2\to2$ annihilation. Moreover, the critical production moment determining the final yield of UV freeze-in locates in a much smaller $x$ (and thus much higher temperature) than the IR freeze-in case.

\subsection{Limits from BBN}\label{sec:BBN}

After EWSB, the SM-like Higgs doublet needs to be replaced by:
\begin{eqnarray} 
H = \left(\begin{array}{c} G^+ \\  \frac{1}{\sqrt{2}}(v + h + i G^0 )  \end{array}\right)~,
\end{eqnarray} 
where $v=246$ GeV is the VEV of SM Higgs~\footnote{If wino decays much later than electroweak phase transition, then $v=246$ GeV is a good approximation.} and $h$ is the observed SM-like Higgs scalar.
$G^{\pm}$ ($ G^{-} = (G^{+})^\ast$) and $G^0$ are Goldstone bosons that form the longitudinal modes of SM gauge bosons $W^{\pm}$ and $Z$. As mentioned earlier, the SM-like Higgs VEV will generate mixings among the gauge states $\tilde{B},\tilde{W},\tilde{H}_u,\tilde{H}_d$ and form mass eigenstates of charge-neutral neutralinos $\tilde{\chi}_{1,2,3,4}$ and charged charginos $\tilde{\chi}^\pm_{1,2}$ (with ascending mass order inside sectors of neutralinos and charginos, respectively). For the scenario considered in this work, the component of neutralino $\tilde{\chi}^0_1$ ($\tilde{\chi}^0_2$) is dominated by bino $\tilde{B}$ (wino $\tilde{W}^3$), and component of chargino $\tilde{\chi}^\pm_1$ is dominated by winos $\frac{1}{\sqrt{2}} ( \tilde{W}^1 \mp i \tilde{W}^2 )$. More details of the approximated masses and couplings can be found in \cite{Gunion:1987yh,Gunion:1992tq,Djouadi:2001fa}. In the following, we would utilize the language of gauge states (bino $\tilde{B}$, wino $\tilde{W}$, higgsinos $\tilde{H}_u,\tilde{H}_d$ ) and mass eigenstates (neutralino $\tilde{\chi}^0$, chargino $\tilde{\chi}^\pm$) interchangeably before and after EWSB.

Now we study the limit of BBN on our scenario from lifetimes of neutralinos, charginos.
In our scenario, only neutralino $\tilde{\chi}^0_1 \approx \tilde{B},\  \tilde{\chi}^0_2 \approx \tilde{W}^3$ and chargino $\tilde{\chi}^\pm_1 \approx \frac{1}{\sqrt{2}} ( \tilde{W}^1 \mp i \tilde{W}^2)$ existed in the primordial thermal bath. 
Due to the loop induced mass-splitting between $\tilde{\chi}^\pm_1$ and $\tilde{\chi}^0_2$, chargino
$\tilde{\chi}^\pm_1$ can have the 2-body decay  $\tilde{\chi}^\pm_1 \to \tilde{\chi}^0_2 \pi^{\pm}$~\cite{Yamada:2009ve,Ibe:2012sx,McKay:2017xlc, Ibe:2022lkl}. 
It makes the lifetime of $\tilde{\chi}^\pm_1$ much shorter than 1 sec, and thus not affecting the BBN profile.
However, we need to scrutinize the lifetime of $\tilde{\chi}^0_2$ more carefully. If $\tilde{\chi}^0_2$ decays after the onset of BBN, then the highly energetic decay products will cause the photodissociation or hadrodissociation and thus change the final abundances of light elements. So a bound from BBN can be put on the model parameters, especially on the SUSY scale $M_{\text{SUSY}}$ ~\cite{Kawasaki:2008qe,Kawasaki:2017bqm}.  

It is easy to see that Fig.~\ref{fig-schematic-process} implies the 2-body decay mode of $\tilde{\chi}^0_2 \to \tilde{\chi}^0_1  h$ at the level of dim-5 after EWSB, in which case we will have:
\begin{eqnarray} 
\mathcal{L}_{\text{eff}} &=&  - \sum_{b=1}^{3} \frac{ (\sqrt{2} g_1 Y_H) (\sqrt{2} g_2  ) }{\mu} \sin\beta \cos\beta (H^\ast  \frac{1}{2}\sigma^b  H) (\tilde{B}\tilde{W}^b)  + h.c.  \nonumber\\
&=&  -  \frac{ g_1 g_2 v }{2 \mu} \sin\beta \cos\beta   ( G^{\mp} \tilde{W}^{\pm} \tilde{B}  - h \tilde{W}^{0} \tilde{B} + h.c.  )  \nonumber\\
&\approx&  -  \frac{ g_1 g_2 v }{2 \mu} \sin\beta \cos\beta   ( G^{\mp} \tilde{\chi}^\pm_1 \tilde{\chi}^0_1  - h \tilde{\chi}^0_2 \tilde{\chi}^0_1 + h.c.  )~,
\label{Leff-2-body-decay}
\end{eqnarray} 
where the first term containing Goldstone boson $G^{\mp}$ can be understood in the context of Goldstone equivalence theorem (GET) for $\tilde{\chi}^\pm_1 \to \tilde{\chi}^0_1 W^{\pm}$.
It should be noticed that Eq.(\ref{Leff-2-body-decay}) does not contain the three-particle coupling $G^0\tilde{W}\tilde{B}$ and thus would not provide a way of inferring the 2-body decay mode $\tilde{\chi}^0_2 \to \tilde{\chi}^0_1 Z$ via the GET. In fact,  $\tilde{\chi}^0_2 \to \tilde{\chi}^0_1 \ Z$ comes from the gauge covariant kinetic terms of gauginos and higgsinos combined with gaugino mixings after EWSB. However, the decay width of $\tilde{\chi}^0_2 \to \tilde{\chi}^0_1  Z$ suffers from an extra suppression of $\frac{1}{\mu^2}$ embedded in the mass mixings compared to $\tilde{\chi}^0_2 \to \tilde{\chi}^0_1 h$ and thus can be ignored \cite{Rolbiecki:2015gsa} . Therefore, we have the following dominant 2-body decay (see Appendix \ref{Appendix-F} for more details):
\begin{eqnarray} 
\Gamma_{\tilde{\chi}^0_2 \to \tilde{\chi}^0_1 h} \approx M_2 \frac{1}{16\pi}  \left(  \frac{ v }{\mu} g_1 g_2\sin\beta \cos\beta \right)^2 \left( 1 - \frac{M_1^2}{M_2^2}  \right) \left(1+\frac{M_1}{M_2} \right)^2~.
\end{eqnarray} 
Using the GET we would obtain the same results for $\Gamma_{\tilde{\chi}^\pm_1 \to \tilde{\chi}^0_1 W^{\pm}}$ when neglecting the gauge boson masses. 

\begin{figure}[ht]
  \centering
  \includegraphics[width=9cm]{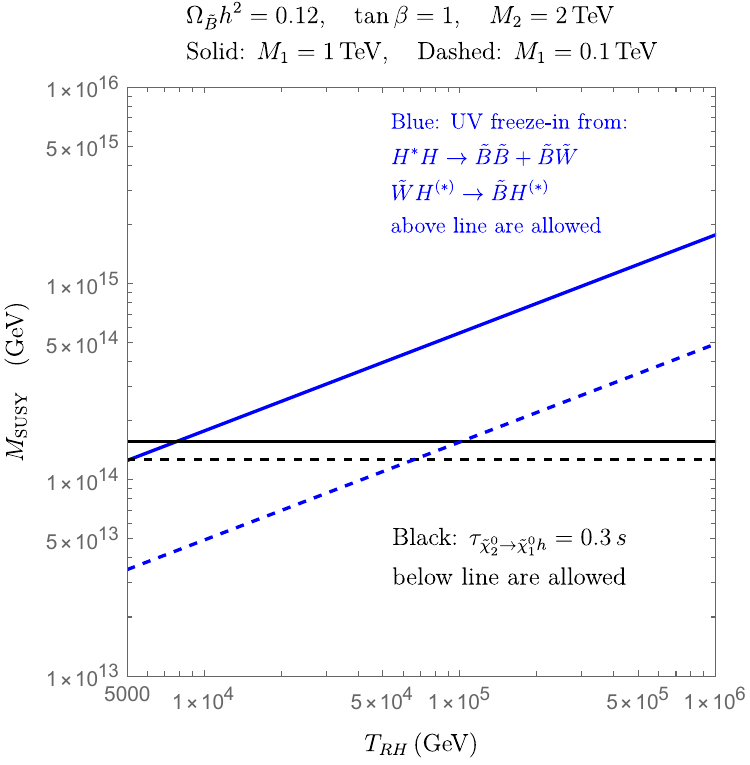} \hspace{1 cm}
  \caption{Interplay between BBN constraints and freeze-in production, where regions below black lines are allowed while region above blue lines are allowed.}
  \label{fig-04}
\end{figure}

In this work, we apply the limit of BBN to the requirement that lifetime of $\tilde{\chi}^0_2$ must be less than 0.3 second \cite{Hall:2009bx}. 
In Fig.~\ref{fig-04} , we show the interplay between BBN constraints and freeze-in production, where regions below black lines are allowed while region above blue lines are allowed. We can see that for bino mass around 0.1-1 TeV, an upper bound of $M_\text{SUSY}\sim 10^{14}$ TeV is needed to satisfy both phenomenological requirements.

\subsection{Limits from direct/indirect detection}\label{sec:dd}

Our scenario can easily escape from the current limits from the direct and indirect detection. In the case of direct detection, Eq.(\ref{Leff_HHBB}) after EWSB would generate the $t$-channel scattering of $\tilde{\chi}^0_1$ with quarks and gluons in SM neucleons mediated by SM Higgs, of which the event rate is suppressed by $1/\mu^2$ and thus negligibly small. In the case of indirect detection, which is basically the inversed process of the freeze-in DM production, would generate cosmic rays via DM pair annihilations $\tilde{\chi}^0_1 \tilde{\chi}^0_1 \to h^\ast \to \text{SM}$ and $\tilde{\chi}^0_1 \tilde{\chi}^0_1 \to h h \to \text{SM}$, of which the flux is again suppressed by $1/\mu^2$ and thus not violating the current experimental bounds.

\subsection{Limits from the LHC}\label{sec:collider}

The collider signals of our scenario mainly come from $p p \to \tilde{\chi}^\pm_1 \tilde{\chi}^\mp_1, \tilde{\chi}^\pm_1 \tilde{\chi}^0_2$ followed by $\tilde{\chi}^\pm_1 \to \tilde{\chi}^0_2 \pi^\pm$ and $\tilde{\chi}^0_2 \to \tilde{\chi}^0_1 h$ 
which both generate the long-lived particle (LLP) signals. The LLP signatures manifest as disappearing track for $\tilde{\chi}^\pm_1 \to \tilde{\chi}^0_2 \pi^\pm$ and displaced vertices for $\tilde{\chi}^0_2 \to \tilde{\chi}^0_1 h$, respectively. However, $\tau_{\tilde{\chi}^0_2 \to \tilde{\chi}^0_1 h} > \mathcal{O}(10^{-2}) \, \text{s}$ would make $\tilde{\chi}^0_2$ traverse through the whole detector before decaying without leaving any energy deposit in the calorimeters, thus can easily evade the current ATLAS \cite{ATLAS:2022zhj} and CMS \cite{CMS:2022qej} searches for displaced vertex signals at $\sqrt{s}= 13\, \text{TeV}$. As for the disappearing track signature of $\tilde{\chi}^\pm_1 \to \tilde{\chi}^0_2 \pi^\pm$, ATLAS \cite{ATLAS:2022rme} and CMS \cite{CMS:2020atg} also performed dedicated searches using dataset at $\sqrt{s}= 13\, \text{TeV}$ and imply that $\tilde{\chi}^\pm_1, \tilde{\chi}^0_2$ should be heavier than 500-600 GeV, therefore our benchmark points with $M_2 = 2 \, \text{TeV}$ are still available.

\subsection{Discussions}\label{sec-discussion}

Before ending this section, we discuss some details concerning the SUSY mass spectrum. 

Firstly, our findings indicate that, to achieve the correct dark matter relic abundance through the UV freeze-in mechanism, the typical mass scales of SUSY particles (excluding gauginos) should be in the range of $10^{13-14}$ GeV. An intriguing question arises concerning whether the SM Higgs with mass around 125 GeV can be accommodated within this framework. In heavy SUSY scenarios, as discussed in \cite{Giudice:2011cg, Hall:2014vga}, sparticle masses around $10^{13}$ GeV are still viable, particularly when considering $\tan\beta = 1$ and allowing for the uncertainty in SM parameters within  $1\sigma$ range. Expanding the range of uncertainty in SM parameters, particularly the top Yukawa coupling, to $2\sigma$ range allows for a significant upward adjustment of the SUSY mass scale. Notably, the work of \cite{Ellis:2017erg} delves into high-scale SUSY within  $3\sigma$ uncertainty for Standard Model parameters, with findings indicating that for $\tan\beta=1$ a SUSY scale as high as $10^{16}$ GeV remains consistent with the observed SM Higgs mass. This underscores the importance of considering a reasonable range of uncertainty in SM parameters when assessing SUSY scenarios. However, it is crucial to note that future precision measurements of SM parameters hold the potential to rigorously scrutinize SUSY scenarios. Therefore, our model stands poised for being tested against these precise measurements, providing an avenue for further validation and refinement.

Secondly, in our study we adopted the assumption of a gluino mass greater than the reheating temperature $T_{RH}$ to avoid potential conflict with BBN constraints. Concurrently, we  presented typical mass ranges for the bino (and wino) falling within the span of 0.1-1 TeV, with $T_{RH}$ estimated at around $10^4 - 10^6$ GeV. This naturally entails the requirement for a substantial hierarchy between the gluino mass and the bino (as well as the wino) mass. Achieving such a hierarchy within the domain of supersymmetry calls for a meticulous consideration of the scenarios associated with SUSY breaking and mediation. One plausible avenue involves postulating non-universal gaugino masses. This can be accomplished by ascribing distinct representations to the SUSY breaking superfield $\Phi$ with non-vanishing F-terms (see e.g. \cite{Martin:2009ad,Martin:2013aha,Raza:2018jnh}). While the above framework provides a well-recognized means of introducing a phenomenologically oriented hierarchy among gaugino masses, attaining the desired mass ratio between the gluino and the bino/wino may necessitate a fine-tuning of the contributions arising from these different representations. 

It is crucial to emphasize that our present research predominantly focuses on delving into the phenomenological aspects, especially within the realm of dark matter. Acknowledging that a comprehensive model incorporating precise calculations of the Higgs mass and the requisite mass hierarchy for gauginos is undoubtedly imperative, we intend to actively explore the feasibility of incorporating these elements in our future work.

\section{Conclusion}\label{sec-conclusion}

We studied a scenario of dark matter generated from UV freeze-in mechanism, realized in the framework of  high scale MSSM. The bino is the dark matter candidate and its relic abundance is generated by the freeze-in processes via the dim-5 or dim-6 operators. We found that the SUSY scale $M_\text{SUSY}$ should be around $10^{13-15} ~\text{GeV}$ for reheating temperature in the range of $10^{4-6}~ \text{GeV}$. We also illustrated the interplay between BBN constraints from neutral wino decay and the experimentally observed dark matter relic abundance, implying an upper bound of $M_\text{SUSY}$ around $10^{14} ~\text{GeV}$ for wino mass around 2 TeV and bino mass of $0.1\sim 1$ TeV.


\addcontentsline{toc}{section}{Acknowledgments}
\acknowledgments
This work was supported by the Natural Science Foundation of China (NSFC) under grant numbers 12105118, 11947118, 12075300, 11821505 and 12335005, the Peng-Huan-Wu Theoretical Physics Innovation Center (12047503), the CAS Center for Excellence in Particle Physics (CCEPP), and the Key Research Program of the Chinese Academy of Sciences under grant No. XDPB15.
CH acknowledges support from the Sun Yat-Sen University Science Foundation and the Fundamental Research Funds for the Central Universities, Sun Yat-sen
University under Grant No. 23qnpy58. PW acknowledges support from Natural Science Foundation of Jiangsu Province (Grant No. BK20210201), Fundamental Research Funds for the Central Universities, Excellent Scholar Project of Southeast University (Class A), and the Big Data Computing Center of Southeast University.
\vspace{0.5cm}

\appendix
\section{Notation conventions and dim-5 operator in Case I}
\label{Appendix-A}

In Eq.(\ref{eq-L-master}), the dot product means $\tilde{H}_u \cdot \tilde{H}_d = \tilde{H}_{u,i} (i\sigma^2)^{ij} \tilde{H}_{d,j} = \tilde{H}^+_u \tilde{H}^-_d - \tilde{H}^0_u \tilde{H}^0_d $
to realize the isospin symmetry ${\rm SU(2)_L}$ where $\sigma^2$ is the second Pauli matrix.
The Kronecker delta function $\delta_i^{\,\,\,j}$ manifests the ${\rm SU(2)_L}$-blindness of the ${\rm U(1)_Y}$ interactions under consideration for binos production
and $Y_{H_u}=+1/2,Y_{H_d}=-1/2$ are the hypercharges of doublets $H_u, H_d$, respectively. We follow the convention of \cite{Dreiner:2008tw} and  impose the left-chiral two-component spinor formalism for  higgisnos $\tilde{H}^+_u, \tilde{H}^0_u, \tilde{H}^0_d, \tilde{H}^-_d$ and bino $\tilde{B}$ (as well as winos $\tilde{W}$ and gluinos $\tilde{g}$ in later discussion). 
For the Case I in Section \ref{bino-freezein}, the relevant Lagrangian terms are
 \small 
\begin{eqnarray} 
\mathcal{L} &\supset& - \frac{1}{2}  M_1 \tilde{B} \tilde{B}  -   \mu \left(  \tilde{H}^+_u \tilde{H}^-_d - \tilde{H}^0_u \tilde{H}^0_d   \right)  + h.c. \nonumber\\
 & & - \frac{g_1}{\sqrt{2}} ( {H}^+_u )^\ast  \tilde{H}^+_u  \tilde{B} - \frac{g_1}{\sqrt{2}} ( {H}^0_u )^\ast  \tilde{H}^0_u  \tilde{B} + \frac{g_1}{\sqrt{2}} ( {H}^-_d )^\ast  \tilde{H}^-_d  \tilde{B} + \frac{g_1}{\sqrt{2}} ( {H}^0_d )^\ast  \tilde{H}^0_d  \tilde{B} + h.c. ~,
\end{eqnarray} 
\normalsize
After integrating out higgsinos with mass $\mu$, we obtain dim-5 operator between SM Higgs $H_\text{SM}$ and $\tilde{B}$ DM:
\begin{eqnarray} 
\mathcal{L}^{\text{eff}}_{HH^\ast \to \tilde{B}\tilde{B}} 
&=&  -\frac{(\sqrt{2} g_1 Y_H)(\sqrt{2} g_1 Y_H)}{\mu} ( H_u^\ast \cdot H_d^\ast ) \tilde{B} \tilde{B} + h.c. \nonumber \\
&=& -\frac{2 g_1^2 \, Y_H^2}{\mu} \sin\beta  \cos\beta  (  H^\ast_{\rm SM} \cdot i \sigma^2 H_{\rm SM} ) ( \tilde{B} \tilde{B} + \tilde{B}^{\dagger} \tilde{B}^{\dagger} ) \nonumber \\
&=& \frac{2 g_1^2 \, Y_H^2}{\mu} \sin\beta  \cos\beta  ( | H_{\rm SM} |^2 )  ( \tilde{B} \tilde{B} + \tilde{B}^{\dagger} \tilde{B}^{\dagger} )~,
\label{Leff_HHBB_full}
\end{eqnarray} 
where $Y_H=|Y_{H_u}|=|Y_{H_d}|=1/2$ and the dot products are $H_u^\ast \cdot H_d^\ast = ({H}^+_u)^\ast ({H}^-_d)^\ast - ({H}^0_u)^\ast ({H}^0_d)^\ast$.

\section{Boltzmann equation and calculation details of freeze-in DM in Case I \label{Appendix-B} }
In the homogeneous and isotropic universe, the production of bino is described by following Boltzmann equation \cite{Kolb:1990vq}:
\begin{eqnarray}
\frac{d}{d t} n_{\tilde{B}} + 3 \mathcal{H} n_{\tilde{B}} = \textbf{C} ~,
\end{eqnarray}
with $n_{\tilde{B}}$ denoting the number density of bino particle, and $\mathcal{H}$ is the Hubble expansion rate. Taking $ H H^\ast \to {\color{black} \tilde{B} \tilde{B}} $ ($\tilde{B}$ means the physical bino particle) in Case I of Section \ref{bino-freezein} as an example, we have \cite{Edsjo:1997bg}
\begin{eqnarray}
\label{collision-HHBB-master-formula}
\textbf{C}_{ij\to kl} &=& N \times \frac{1}{S}  \times \bigg\{ \int \frac{d^3 p_i}{ (2\pi)^3 2E_i} \frac{d^3 p_j}{ (2\pi)^3 2E_j} \frac{d^3 p_k}{ (2\pi)^3 2E_k} \frac{d^3 p_l}{ (2\pi)^3 2E_l}  \nonumber \\
& &  \times   (2\pi)^4\delta^4(p_i+p_j-p_k-p_l)  \ \bigg[ f_i f_j ( 1- f_k)(1- f_l) - f_k f_l (1+ f_i)(1+ f_j)  \bigg] \nonumber \\
& &  \times \sum_{\rm internal \, d.o.f} |\mathcal{M}|^2_{ij \to kl} \ \bigg\}~,
\end{eqnarray}
where $f_{i,j,k,l}$ are the phase space distribution functions. The number density, taking $f_i$ as example, is defined as
\begin{eqnarray}
n_i \equiv g_i \int \frac{d^3 p}{(2 \pi)^3} f_i (p) ~,
\end{eqnarray}
in which $g_i$ is the internal degree of freedom (d.o.f.) of particle $i$.
{\color{black} The factor $N$ denotes the number of particles under consideration produced in the final state and the factor $1/S$ originates from the phase space suppression due to the \textit{identical} particles in the initial and final states}. For $ H H^\ast \to {\color{black} \tilde{B} \tilde{B}} $ we have $N=2$ and $1/S=1/(N!)=1/2$.
After some manipulations and neglecting the negligible backward process, we have \cite{Edsjo:1997bg}
\begin{eqnarray}
\label{collision-HHBB-explicit-phase-space-integral}
\textbf{C}_{ij \to kl} &\approx& \frac{T}{32\pi^4} \int^{\infty}_{(m_k + m_l)^2} ds \, p_{ij} \, W_{ij \to kl} \, K_1( {\sqrt{s}}/{T})  \\ 
W_{ij \to kl} &=& \frac{p_{kl}}{16 \pi^2 \sqrt{s}}  \,  \sum_{\rm internal \, d.o.f}  \int\, d\Omega \, |\mathcal{M}|^2_{ij \to kl} \\ 
p_{ij} &=&  \frac{\sqrt{s-(m_i+m_j)^2}\sqrt{s-(m_i-m_j)^2}}{2\sqrt{s}} ~,
\end{eqnarray}
where $p_{kl}$ is similar to $p_{ij}$. After summing over all bino spin states $s_1,s_2$ and isospin states of the SM-like Higgs , we have the amplitude square ($s$ is the square of the central energy):
\begin{eqnarray}
&&\sum_{\rm internal \, d.o.f}  \int d\Omega \  |\mathcal{M}|^2_{ H H^\ast \to  \tilde{B} \tilde{B} } ~,\nonumber \\
&&\approx   (2\pi)  \times \bigg[ \sum_{i,j=1}^{2} (\delta_i^{\,\,\,j})^2 \bigg] \bigg[  Y^4_{H} \bigg]  \bigg(  \frac{ g_1 g_2 \sin\beta \cos\beta }{\mu} \bigg)^2 \bigg[ 64 \ s \bigg( 1 - \frac{4 M_1^2}{s}  \bigg)^{\frac{3}{2}} \bigg]  \nonumber \\
&&\approx (16\pi) \times \frac{g_1^4}{\mu^2} \sin^2\beta  \cos^2\beta \ s~.
\end{eqnarray}
We modify the MSSM model file available in {\bf FeynRules} \cite{Christensen:2008py,Alloul:2013bka} to highlight the gauge state interactions and then export to {\bf FeynArts} \cite{Hahn:2000kx} augmented with {\bf FeynCalc} \cite{Shtabovenko:2016sxi} to perform the calculation.

Since we are considering freeze-in production of $\tilde{B}$, $f_{1,2}$ in Eq.~(\ref{collision-HHBB-master-formula}) can be ignored. 
We can further approximate $f_{3,4}$ by Maxwell-Boltzmann distribution, i.e. $f_{3,4} \approx e^{-E_{3,4}/T}$.
Then the collision term can be rewritten as~\cite{Edsjo:1997bg,Gondolo:1990dk,Elahi:2014fsa}
\begin{eqnarray}
\label{collision-HHBB-final-form}
\textbf{C}_{HH^\ast \to \tilde{B}\tilde{B}} & \approx & 
\frac{T}{ 2048 \pi^6 } \int^{\infty}_{4M_1^2} ds \  ( s-4M_1^2 )^{1/2} K_1( {\sqrt{s}}/{T})  \sum_{\rm internal \, d.o.f}  \int\, d\Omega \, |\mathcal{M}|^2_{HH^\ast \to \tilde{B}\tilde{B}} ~, \nonumber \\
& \approx &  \frac{T}{ 128 \pi^5 } \frac{ g_1^4\, \sin^2\beta  \cos^2\beta }{\mu^2 }  \int^{\infty}_{4M_1^2} ds \  s^{3/2} K_1( {\sqrt{s}}/{T})~.
\end{eqnarray}
Here $K_1$ is the Bessel function of the second kind, and we treat the SM-like Higgs in the initial state as being massless. 
In the case where $M_1 \ll T$, the collision term can be approximated as (using {\color{black}  $\int^{\infty}_0 dx x^4 K_1(x) = 16$})
\begin{eqnarray}
\int^{\infty}_{4M_1^2} ds \  s^{3/2} K_1( {\sqrt{s}}/{T})
&\approx& \int^{\infty}_{0} (dx\, T) \ (2xT) (xT)^3 K_1(x)\nonumber\\ 
&=& 2\,T^5 \int^{\infty}_0 dx \, x^4 K_1(x) = 32 \ T^5~.
\end{eqnarray}

\section{The calculation details in Case II \label{Appendix-C} }
We use $f=q,l$ with $q = u_L, d_L , u^\dagger_R, d^\dagger_R $ and $l = \nu, e_L, e^\dagger_R$ to denote the left-handed two-component Weyl spinor of SM quarks and leptons, where the bars are simply notations and do not mean the Dirac conjugation. 
Hypercharges are given by $\{Y_{Q_L}=Y_{u_L}=Y_{d_L},\ Y_{{u^\dagger_R}},\ Y_{{d^\dagger_R}},\ Y_{L_L}=Y_{e_L}=Y_{\nu},\ Y_{{e^\dagger_R}}  \} = \{1/6,\ -2/3,\ 1/3,\ -1/2,\ 1  \}$. 
After integrating out sfermions with mass $M_{\tilde{f}}$ in the right panel of Fig.\ref{fig-schematic-process}, we obtain dim-6 operators between SM fermion pair and $\tilde{B}$ pair:
\begin{eqnarray} 
\mathcal{L}_{\text{eff}} =  \sum_{f=q,l} \frac{ (\sqrt{2} g_1 Y_f)(\sqrt{2} g_1 Y_f)}{M^2_{\tilde{f}}} ( f^{\dagger} \tilde{B}^\dagger ) (f \tilde{B}) ~,
\end{eqnarray} 
where for simplicity we consider an universal mass for all the fermions, i.e. $M_{\tilde{f}} \equiv M_{\tilde{q}} = M_{\tilde{l}}$.

The amplitude squared terms in the collision term for $f \bar{f} \to \tilde{B} \tilde{B}$ scattering process is given by\footnote{
Again, fields in the initial and final states in the process should be understood in the sense of physical particles, where $\bar{f}$ denotes the physical anti-particle. Discussion on the naming convention of particles, states and filed can be found in, e.g. \cite{Dreiner:2008tw}.}
\small
\begin{eqnarray} 
\sum_{\rm internal \, d.o.f}  \int d\Omega \  |\mathcal{M}|^2_{f \bar{f} \to \tilde{B} \tilde{B}} 
&\approx& 2\pi N_{\rm flavor}  \bigg[ N_{\rm color} \bigg( \sum_{i,j=1}^{2} (\delta_i^{\,\,\,j})^2  Y^4_{Q_L} + Y_{{u^\dagger_R}}^4 + Y_{{d^\dagger_R}}^4 \bigg)  \nonumber  \\ 
&& + \bigg( \sum_{i,j=1}^{2} (\delta_i^{\,\,\,j})^2  Y^4_{L_L} + Y_{{e^\dagger_R}}^4  \bigg)  \bigg] \bigg(   \frac{g_1^2 }{M^2_{\tilde{f}}}  \bigg)^2 \bigg[ \frac{16}{3} \ s^2  \bigg]  \nonumber\\
&=& \frac{1520\pi}{27}  \frac{g_1^4}{M^4_{\tilde{f}}} \ s^2
\end{eqnarray} 
\normalsize 
where $N_{\rm flavor}=N_{\rm color}=3$. As in Eq.~(\ref{collision-HHBB-master-formula}), if we neglect bino mass, then the collision term can be approximately given by (using {\color{black}  $\int^{\infty}_0 dx x^6 K_1(x) = 384$})
\begin{eqnarray}
\textbf{C}_{f \bar{f} \to \tilde{B} \tilde{B} }  & \approx &  \frac{T}{ 2048 \pi^6 } \int^{\infty}_{4M_1^2} ds \  ( s-4M_1^2 )^{1/2} K_1( {\sqrt{s}}/{T})  \sum_{\rm internal \, d.o.f}  \int\, d\Omega \, |\mathcal{M}|^2_{f f^\dagger \to \tilde{B} \tilde{B}^\dagger }  \nonumber\\
&\approx&  \frac{T}{ 2048 \pi^6 } \bigg( \frac{1520\pi}{27}  \frac{g_1^4}{M^4_{\tilde{f}}}  \bigg) \int^{\infty}_{4M_1^2} ds \  s^{5/2} K_1( {\sqrt{s}}/{T})   \nonumber\\
&\approx&  \frac{T}{ 2048 \pi^6 } \bigg( \frac{1520\pi}{27}  \frac{g_1^4}{M^4_{\tilde{f}}}  \bigg) \int^{\infty}_{0} (T dx) (2Tx)  \  (xT)^{5} K_1( {x})   \nonumber\\
&=& \frac{ 190 }{ 9 } g_1^4  \frac{1}{\pi^5 }  \frac{1}{M^4_{\tilde{f}}}  T^8~.
\end{eqnarray}

\section{The calculation details in Case III A}
\label{Appendix-D}
When neglecting all particle masses in the final state, we have
\begin{eqnarray}
&&\sum_{\rm internal \, d.o.f}  \int d\Omega \   |\mathcal{M}|^2_{ H H^\ast \to  \tilde{B} \tilde{W} }  \nonumber\\ 
&& \approx   (2\pi)  \times \bigg[ \sum_{b=1}^{3} \text{tr} \bigg( \frac{1}{2} \sigma^b \frac{1}{2} \sigma^{b} \bigg) \bigg] \bigg[  Y^2_{H} \bigg] \bigg(  \frac{ g_1 g_2 \sin\beta \cos\beta }{\mu}  \bigg)^2 \bigg[ 64 \ s \bigg]   \nonumber\\
&&= (48\pi)  \times \frac{g_1^2 g_2^2}{\mu^2} \sin^2\beta  \cos^2\beta \ s ~, \\ \nonumber\\
&&\sum_{\rm internal \, d.o.f}  \int d\Omega \   |\mathcal{M}|^2_{ \tilde{W} H \to \tilde{B} H } =  \sum_{\rm internal \, d.o.f}  \int d\Omega \ |\mathcal{M}|^2_{ \tilde{W} H^\ast \to \tilde{B} H^\ast }   \nonumber \\ 
&& \approx   (2\pi)  \times  \bigg[ \sum_{b=1}^{3} \text{tr} \bigg( \frac{1}{2} \sigma^b \frac{1}{2} \sigma^{b} \bigg) \bigg] \bigg[  Y^2_{H} \bigg]  \bigg(  \frac{ g_1 g_2 \sin\beta \cos\beta }{\mu}  \bigg)^2 \bigg[32 \ s \bigg]  \nonumber\\
&&= (24\pi)  \times \frac{g_1^2 g_2^2}{\mu^2} \sin^2\beta  \cos^2\beta \ s ~, \\
&&\sum_{\rm internal \, d.o.f}  \int d\Omega \   |\mathcal{M}|^2_{ f \bar{f} \to  \tilde{B} \tilde{W} } \nonumber \\  
&& \approx 2\pi \bigg[ \sum_{b=1}^{3} \text{tr} \bigg( \frac{1}{2} \sigma^b \frac{1}{2} \sigma^{b} \bigg) \bigg]  \bigg[ N_{\rm flavor}  \bigg( Y^2_{L_L} + N_{\rm color}  Y^2_{Q_L}  \bigg)  \bigg]  \bigg(   \frac{g_1 g_2 }{M^2_{\tilde{f}}}  \bigg)^2 \bigg[ \frac{16}{3}   \ s^2  \bigg]  \nonumber\\
&& =(16\pi) \times \frac{ g_1^2 g_2^2 }{ M^4_{\tilde{f}} } \ s^2  ~,\\ \nonumber\\
&&\sum_{\rm internal \, d.o.f}  \int d\Omega \  |\mathcal{M}|^2_{ \tilde{W} f \to  \tilde{B} f } = \sum_{\rm internal \, d.o.f}  \int d\Omega \ |\mathcal{M}|^2_{ \tilde{W} \bar{f} \to  \tilde{B} \bar{f} }   \nonumber \\ 
&& \approx  2\pi \bigg[ \sum_{b=1}^{3} \text{tr} \bigg( \frac{1}{2} \sigma^b \frac{1}{2}  \sigma^{b} \bigg) \bigg]  \bigg[ N_{\rm flavor}  \bigg( Y^2_{L_L} + N_{\rm color}  Y^2_{Q_L}  \bigg)  \bigg]   \bigg(  \frac{g_1 g_2 }{M^2_{\tilde{f}}}  \bigg)^2 \bigg[  \frac{32}{3}  \ s^2  \bigg]  \nonumber \\
&& =  (32\pi) \times \frac{ g_1^2 g_2^2 }{ M^4_{\tilde{f}} } \ s^2 ~, \\
&&\sum_{\rm internal \, d.o.f}  \int d\Omega \   |\mathcal{M}|^2_{ f \bar{f} \to  \tilde{B} \tilde{G} } \nonumber\\
&&\approx 2\pi  \bigg[ \sum_{a=1}^{8} \text{tr} \bigg( \frac{1}{2} \lambda^{a} \frac{1}{2} \lambda^{a} \bigg) \bigg]  \bigg[ N_{\rm flavor} \bigg( \sum_{i,j=1}^{2} (\delta_i^{\,\,\,j})^2 Y^2_{Q_L} + Y_{{u^\dagger_R}}^2 + Y_{{d^\dagger_R}}^2 \bigg) \bigg] \bigg(  \frac{g_1 g_3 }{M^2_{\tilde{f}}}  \bigg)^2 \bigg[ \frac{16}{3}   \ s^2 \bigg]  \nonumber\\
&&=  (\frac{704\pi}{9}) \times  \frac{ g_1^2 g_3^2 }{ M^4_{\tilde{f}} } \ s^2 ~,  \\
&&\sum_{\rm internal \, d.o.f}  \int d\Omega \   |\mathcal{M}|^2_{ \tilde{G} f \to  \tilde{B} f } = \sum_{\rm internal \, d.o.f}  \int d\Omega \ |\mathcal{M}|^2_{ \tilde{G} \bar{f} \to  \tilde{B} \bar{f} }   \nonumber \\ 
&&\approx 2\pi  \bigg[ \sum_{a=1}^{8} \text{tr} \bigg( \frac{1}{2} \lambda^{a} \frac{1}{2} \lambda^{a} \bigg) \bigg]  \bigg[ N_{\rm flavor} \bigg( \sum_{i,j=1}^{2} (\delta_i^{\,\,\,j})^2 Y^2_{Q_L} + Y_{{u^\dagger_R}}^2 + Y_{{d^\dagger_R}}^2 \bigg) \bigg] \bigg(  \frac{g_1 g_3 }{M^2_{\tilde{f}}}  \bigg)^2\bigg[\frac{32}{3} \ s^2 \bigg]  \nonumber\\
&& =  (\frac{1408\pi}{9}) \times  \frac{ g_1^2 g_3^2 }{ M^4_{\tilde{f}} } \ s^2 ~.
\end{eqnarray} 

\section{The calculation details in Case III B}
\label{Appendix-E}

The $1\to3$ decay processes are indicated by the red colored arrow in Fig.\ref{fig-schematic-process}. When neglecting all particle masses in the final state, we have
\begin{eqnarray}
\Gamma_{\tilde{W} \to  \tilde{B} H H^\ast } &=& \frac{1}{(2\pi)^3} \frac{1}{32 M_2^3} \ \frac{1}{g_{\tilde{W}}} \sum_\text{internal d.o.f.} \int dm^2_{12}\ dm^2_{23} \ |\mathcal{M}|^2_{ \tilde{W} \to  \tilde{B} H H^\ast }     \nonumber\\
&=& \frac{1}{(2\pi)^3} \frac{1}{32 M_2^3} \ \frac{1}{ \sum_{b=1}^{3} (2 s_{\tilde{W}} +1) }    
\bigg[ \sum_{b=1}^{3} \text{tr} \bigg( \frac{1}{2} \sigma^b \frac{1}{2} \sigma^{b} \bigg) \bigg] \nonumber\\
&& \times \bigg[  Y^2_{H} \bigg]  \bigg[  \frac{32}{3} M_2^6  \left(  \frac{g_1 g_2 \sin\beta \cos\beta }{\mu} \right)^2 \bigg]  \nonumber\\
&=&\frac{1}{384 \pi^3}  \left(  \frac{ g_1 g_2 \sin\beta \cos\beta }{\mu} \right)^2 M_2^3 ~.\\ \nonumber \\
\Gamma_{\tilde{W} \to \tilde{B}  f\bar{f} } &=& \frac{1}{(2\pi)^3} \frac{1}{32 M_2^3} \ \frac{1}{g_{\tilde{W}}} \sum_\text{internal d.o.f.} \int dm^2_{12}\ dm^2_{23} \ |\mathcal{M}|^2_{ \tilde{W} \to  \tilde{B} f \bar{f} }  \nonumber\\
&=& \frac{1}{(2\pi)^3} \frac{1}{32 M_2^3} \ \frac{1}{ \sum_{b=1}^{3} (2 s_{\tilde{W}} +1) }   
\bigg[ \sum_{b=1}^{3} \text{tr} \bigg( \frac{1}{2} \sigma^b \frac{1}{2} \sigma^{b} \bigg) \bigg]  \nonumber\\
&& \times \bigg[ N_{\rm flavor}  \bigg( Y^2_{L_L} + N_{\rm color}  Y^2_{Q_L}  \bigg)  \bigg]   \bigg[  \frac{2}{3} M_2^8  \left(  \frac{g_1 g_2 }{M^2_{\tilde{f}}} \right)^2 \bigg]  \nonumber\\
&=&\frac{1}{1536 \pi^3}  \left(  \frac{g_1 g_2 }{M^2_{\tilde{f}}} \right)^2 M_2^5 ~, \\  \nonumber \\
\Gamma_{\tilde{G} \to \tilde{B} f\bar{f}  } &=& \frac{1}{(2\pi)^3} \frac{1}{32 M_3^3} \ \frac{1}{g_{\tilde{G}}} \sum_\text{internal d.o.f.} \int dm^2_{12}\ dm^2_{23} \ |\mathcal{M}|^2_{ \tilde{G} \to  \tilde{B} f \bar{f} }     \nonumber\\
&=& \frac{1}{(2\pi)^3} \frac{1}{32 M_3^3} \ \frac{1}{ \sum_{a=1}^{8} (2 s_{\tilde{G}} +1) }  
\bigg[ \sum_{a=1}^{8} \text{tr} \bigg( \frac{1}{2} \lambda^{a} \frac{1}{2} \lambda^{a} \bigg) \bigg] 
 \nonumber\\
&& \times \bigg[ N_{\rm flavor} \bigg(  \sum_{i,j=1}^{2} (\delta_i^{\,\,\,j})^2 Y^2_{Q_L} + Y_{{u^\dagger_R}}^2 + Y_{{d^\dagger_R}}^2 \bigg) \bigg]   \bigg[  \frac{2}{3} M_2^8  \left(  \frac{g_1 g_3 }{M^2_{\tilde{f}}} \right)^2 \bigg]  \nonumber\\
&=&\frac{11}{9216 \pi^3}  \left(  \frac{g_1 g_3 }{M^2_{\tilde{f}}} \right)^2 M_2^5~.
\end{eqnarray}
where $dm^2_{12}, dm^2_{23}$ are defined in \cite{ParticleDataGroup:2022pth}. 

\section{The calculation details of 2-body decay after EWSB}\label{Appendix-F}
As discussed in Section \ref{sec:BBN}, we have the following $1\to2$ decay possibly affecting the cosmological BBN:
\begin{eqnarray} 
\Gamma_{\tilde{\chi}^0_2 \to \tilde{\chi}^0_1 h} &=& \frac{1}{2s_{\Tilde{\chi}^0_2}+1} \frac{1}{2 M_2} \sum_{\text{spin d.o.f.}} \int d\Pi_2 \ |M|^2_{\tilde{\chi}^0_2 \to \tilde{\chi}^0_1 h}  \nonumber\\
&=& \frac{1}{2} \frac{1}{2 M_2} \ \int d\Pi_2 \ \left(  \frac{ v }{\mu} g_1 g_2\sin\beta \cos\beta \right)^2 \bigg[4 (p_{\tilde{\chi}^0_2}\cdot p_{\tilde{\chi}^0_1} + M_{\tilde{\chi}^0_2} M_{\tilde{\chi}^0_1}) \bigg]   \nonumber\\
&\approx& \frac{1}{2} \frac{1}{2 M_2} \ \bigg[\int d\Omega \frac{1}{16\pi^2} \frac{|\Vec{p}_{\tilde{\chi}^0_1}|}{M_2}  \bigg] \ \left(  \frac{ v }{\mu} g_1 g_2\sin\beta \cos\beta \right)^2 \bigg[4 (M_2  E_{\tilde{\chi}^0_1} + M_2 M_1) \bigg]~, \nonumber \\
\end{eqnarray} 
where
\begin{eqnarray} 
E_{\tilde{\chi}^0_1} &=& \frac{M_2^2 + M_1^2 - M_h^2}{2 M_2} \approx \frac{M_2^2 + M_1^2}{2 M_2}~, \\
|\Vec{p}_{\tilde{\chi}^0_1}| &=& \sqrt{E_{\tilde{\chi}^0_1}^2 - M_1^2} = \frac{\left(M_2^4 + M_1^4 + M_h^4 - 2 M_2^2 M_1^2 - 2 M_2^2 M_h^2 - 2 M_1^2 M_h^2 \right)^{\frac{1}{2}} }{2 M_2} \nonumber\\
&\approx& \frac{M_2^2 - M_1^2}{2 M_2}~ .
\end{eqnarray} 
Finally, we have \cite{Gunion:1987yh}
\begin{eqnarray} 
\Gamma_{\tilde{\chi}^0_2 \to \tilde{\chi}^0_1 h} &\approx& \frac{1}{2} \frac{1}{2 M_2} \Bigg[ 4\pi \frac{1}{16\pi^2} \frac{1}{2} \left( 1 - \frac{M_1^2}{M_2^2}  \right) \bigg] \left(  \frac{ v }{\mu} g_1 g_2\sin\beta \cos\beta \right)^2 \bigg[4 M_2 \frac{(M_2 + M_1)^2}{2 M_2} \bigg]  \nonumber\\
&\approx& M_2 \frac{1}{16\pi}  \left(  \frac{ v }{\mu} g_1 g_2\sin\beta \cos\beta \right)^2 \left( 1 - \frac{M_1^2}{M_2^2}  \right) \left(1+\frac{M_1}{M_2} \right)^2~.
\end{eqnarray} 
Using the GET we would obtain the same results in the high energy limit for $\Gamma_{\tilde{\chi}^\pm_1 \to \tilde{\chi}^0_1 W^\pm}$.

\addcontentsline{toc}{section}{References}
\bibliography{ref}
\bibliographystyle{CitationStyle}


\end{document}